\documentclass{article}

\usepackage{arxiv}

\usepackage[utf8]{inputenc} 
\usepackage[T1]{fontenc}    
\usepackage{hyperref}       
\usepackage{url}            
\usepackage{booktabs}       
\usepackage{amsfonts}       
\usepackage{nicefrac}       
\usepackage{microtype}      
\usepackage{lipsum}
\usepackage{graphicx}
\usepackage{amsmath}

\title{Sub-Ensemble Isolation in SERF Magnetometry Enabled by Micrometer-Scale Polarization Control}

\author{
Zihua Liang \\
Institute of Large-scale Scientific Facility and Centre for Zero Magnetic Field Science, \\
Beihang University, \\Beijing 100191, China.
   \And
Yuhao Zhang \\
Institute of Large-scale Scientific Facility and Centre for Zero Magnetic Field Science, \\
Beihang University, \\Beijing 100191, China.
   \And
Lu Liu \\
Institute of Large-scale Scientific Facility and Centre for Zero Magnetic Field Science, \\
Beihang University, \\Beijing 100191, China.
     \And
Jinsheng Hu \\
Institute of Large-scale Scientific Facility and Centre for Zero Magnetic Field Science, \\
Beihang University, \\Beijing 100191, China.
     \And
Peng Zhou\\
Institute of Large-scale Scientific Facility and Centre for Zero Magnetic Field Science, \\
Beihang University, \\Beijing 100191, China.
     \And
Gen Hu\\
Institute of Large-scale Scientific Facility and Centre for Zero Magnetic Field Science, \\
Beihang University, \\Beijing 100191, China.
     \And
Gaopu Hou \\
Institute of Large-scale Scientific Facility and Centre for Zero Magnetic Field Science, \\
Beihang University, \\Beijing 100191, China.
     \And
Mao Ye* \\
Institute of Large-scale Scientific Facility and Centre for Zero Magnetic Field Science, \\
Beihang University, \\Beijing 100191, China.
  \texttt{maoye@buaa.edu.cn} \\
}

\begin{document}
\maketitle

\begin{abstract}
Conventional understanding of spin-exchange relaxation-free (SERF) atom ensemble pertains to the common perception that the rapid exchange of atom state finally results in uniform time evolution of the whole ensemble. However, in this study, we demonstrate that by manipulation of pumping polarization in micro-meter level, misalignment between the time evolution of different sub-ensemble can be created within single SERF ensemble with unprecedent independency. A novel pumping system consists of a miniaturized $^{87}$Rb vapor cell and a space-variant polarization metasurface is developed for the prove of concept. Our method induces position-dependent atomic anisotropy in both pumping and absorption into the thermal atomic ensemble. By constructing calculated Zeeman-sublevel populations in SERF regime, distinct sensing channels are generated with 0.22 $V\ nT^{-1}$ average scale factor, which is comparable with single channel generated by single SERF ensemble. Average crosstalk ratio between adjacent channels (between micron scale sub-ensembles) are measured up to 32 dB, through the excitation of fictitious magnetic field (3.5 nT, 30 Hz), which is measured as 20 dB without sub-ensemble isolation in same experimental condition. Our work demonstrates unprecedented spatial resolution in SERF magnetometry which hold new promises for applications including high-spatial resolution neural biomagnetism mapping and portable magnetism measurement device.
\end{abstract} 


\section{Introduction}
The conventional understanding of spin-exchange relaxation-free (SERF) atomic ensembles hold that rapid spin-exchange interactions lead to uniform time evolution of the entire ensemble, thereby prolonging the relaxation time of alkali atoms and enabling ultra-high-sensitivity magnetic field measurements\cite{kominis2003subfemtotesla,dang2010ultrahigh}; however, this uniformity also limits independent manipulation and interrogation of spin polarization, limiting applicability in sub-millimeter high-resolution biomagnetism imaging\cite{aslam2023quantum,boto2018,gilmore2018imaging}. Although additional reductions in size, weight and power can be achieved from further scaling down the numerous individual pieces that comprise an atomic magnetometer, limited by the trade-off between sensitivity and probe size, extant spatial resolution of arrayed magnetometer remain at the centimeter-scale\cite{boto2018,wyllie2012,yuan2023compact,limes2020portable,tierney2019optically,coussens2024modular,zhang2025design}. Moreover, despite DMD- and AOM-based dynamic light-field modulation enables virtual channel partitioning in submillimeter scale\cite{lu2022scanning,fang2020high,dong2019spin}, their inherent optoelectronic complexity fundamentally impedes chip-scale integration and wider application in biomagnetism application. In addition, emerging single cell multi-channel magnetometers offer a compelling solution to address the miniaturization and high-resolution required to enable sub-millimeter pixel imaging in biomagnetism measurement, but brings with them a new set of challenges in reducing magnetic field crosstalk between sensing channels caused by atomic diffusion\cite{ma2025adaptive,vengalattore2007high}. Recent studies demonstrate that inter-channel crosstalk suppression can be achieved by physically separating detection channels beyond the atomic diffusion length\cite{li2024ultra,zhang2023fast,colombo2016four,iivanainen2024four}. However, such approaches inherently involve critical drawbacks: sophisticated optical configurations are required which result in large system volume and signal attenuation caused by multi-layer beam routing. These problems become particularly severe when scaling to high-density arrays in biomagnetism applications, needless to say those conventional isolation methods are not applicable to emerging microfabricated vapor cells and CMOS compatibility\cite{mottola2023optical,raghavan2024functionalized}. 

The recent advancements in photonics and nanofabrication lead to the emergence of wavefront-shaping platforms based on subwavelength-spaced arrays of optical elements (metasurfaces)\cite{dorrah2021metasurface,yu2011light,arbabi2023advances,dorrah2022tunable}, which have received widespread attention in integrated nanophotonic chip-scale alkali vapor magnetometer due to their exquisite beam manipulation capabilities and scalability for compact device integration\cite{xusilicon,hu2024chipACS,yang2024atomic,liang2024metasurface}. Besides, integrated photonics offers the possibility of dramatically miniaturizing sophisticated system functionality to the size scale of a semiconductor chip, including the light delivery and collection processes vital to atom interrogation in a SERF magnetometer\cite{liang2025space,hu2024integrated,li2024high,hu2024chip}.
In this study, we demonstrate that micrometer-scale manipulation of the pumping polarization can induce controlled misalignment in the time evolution of distinct sub-ensembles within a single SERF ensemble. By inducing position-dependent atomic anisotropy in both pumping and absorption into the $^{87}$Rb thermal atomic ensemble, this polarization-imposed channel isolation (PICI) pumping scheme creating spatially separated detection channels with antiparallel macroscopic spin polarization. Theoretical modeling reveals that the polarization gradient introduced by the PICI pumping scheme reduces the relaxation distance by 46$\%$ compared with the uniform pumping scheme based on a quarter-wave plate (QWP). For the proof of concept, a  $2\times2\ mm^2$  space-variant polarization conversion metasurface is designed, fabricated and finally characterized through stokes measurement. Combining this metasurface with SERF magnetometry systems, a multichannel magnetometer system is established with averaged scale factor of 0.22 $V\ nT^{-1}$. To experimentally validate the spatial resolution of this system over conventional uniformly polarized light pumping, a localized fictitious excitation method is introduced, enabling precise micron-scale excitation of specific sub-regions within the SERF ensemble. Crosstalk is quantified by comparing the response signal ratios of adjacent channels under identical local magnetic field excitation across the two pumping schemes. Compared with conventional uniformly polarized pumping scheme that achieves 20 dB average crosstalk ratio, the PICI pumping scheme demonstrates up to 32 dB average crosstalk ratio between adjacent sensing channels. This 12 dB average enhancement corresponding to a 74.9$\%$  average crosstalk reduction and promising three times spatial resolution improvement. With this micro-meter level pumping polarization, the PICI pumping scheme provides a promising glimpse into future compact, high-sensitivity, and high-resolution biomagnetism imaging and pave the way for the advanced portable atomic sensing applications.

\section{Methods}

\subsection{Principle of the PICI Pumping Scheme}

\begin{figure}[ht!]
  \includegraphics[width=\linewidth]{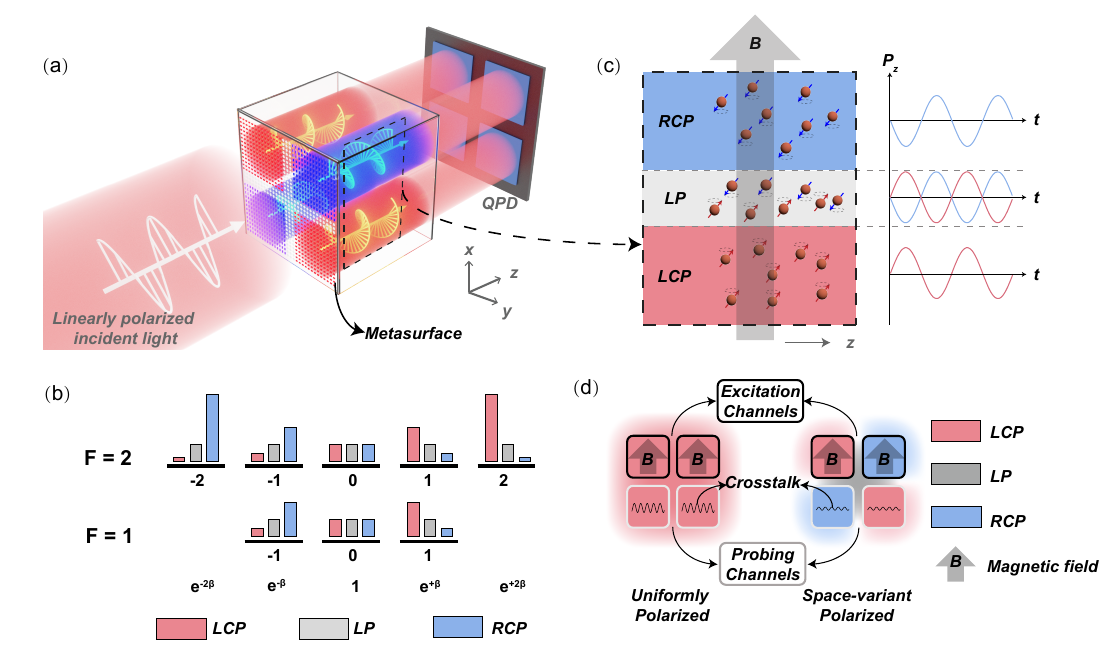}
  \caption{Concept of polarization-imposed channel isolation (PICI). (a) Schematic of multichannel magnetometers with PICI. The spirals within the vapor cells indicate left-/right- handed circular polarization; the red and blue regions indicate electron spin with opposite direction. (b) Zeeman sublevel populations with different polarized pumping beam in SERF regime. LCP, left-handed circular polarization; LP, linear polarization; RCP, right-handed circular polarization. (c) Electron spin in the PICI pumping scheme. B, magnetic field; $P_z$, electron spin polarization. (d) Schematic of the crosstalk for traditional uniformly polarized pumping scheme and PICI pumping scheme. }
  \label{fig:fig1}
\end{figure}

In contrast to conventional single-cell multi-channel magnetometers, where optical polarization transformed by the bulky optics are controlled to uniformly transfer same photon spin to the whole atomic ensemble, magnetometers based on polarization-imposed channel isolation (PICI) pumping scheme utilizes beams with structured polarization to manipulate photon spin point by point at the nanoscale across the transverse plane, thereby creating antiparallel macroscopic spin polarization and naturally separated sensing channels as shown in \textbf{Figure \ref{fig:fig1}(a)}. This versatile polarization response is designed to generated space-variant polarization when incident by linear polarized light: left-/right- handed circular polarizations in four distinct area surrounded by linear polarizations along their edges—equivalent to sequential interactions with multiple polarizers and waveplates.

Pumped by the aforementioned space-variant polarized beam, adjacent atomic sub-ensembles are pumped into distinct Zeeman sublevels (as shown in \textbf{Figure \ref{fig:fig1}(b)}) through photon spin transfer, creating antiparallel macroscopic spin polarization between adjacent circular polarized pumping channels as shown in \textbf{Figure \ref{fig:fig1}(c)} \cite{tierney2019optically,liang2025four}. Influenced by the magnetic field B, the electron spin in these pumping channels generate regular precession signals $P_z$ with a 180-degree relative phase difference. Moreover, due to the presence of equally distributed left- and right- handed circular polarization, the total electron spin $P_z$ in the LP (linear polarization) pumping region undergoes mutual cancellation, thereby insensitive to the variant in magnetic field. This natural characteristic makes it perfectly separating adjacent sensing channels and reducing signal crosstalk caused by thermal diffusion. Furthermore, higher-order photon angular momentum mode in the PICI pumping scheme induces corresponding electron spin polarization modes. These polarization modes exhibit faster relaxation process due to inhomogeneity in electron spin compared to fundamental modes\cite{wang2024mode,shaham2020quantum}, thereby reducing relaxation distance and crosstalk between channels, which is detailed in Theoretical Analysis section. In addition, due to the position-dependent atomic absorption anisotropy, linearly polarized components are completely absorbed by the vapor cell while circular polarized components remained, physically separating four sensing channels in the transmitted light as shown in \textbf{Figure \ref{fig:fig1}(a)}\cite{seltzer2008developments}.

To experimentally validate the spatial resolution of PICI pumping scheme over conventional uniformly polarized light pumping, a localized fictitious magnetic field is used to excite specific regions as shown in \textbf{Figure \ref{fig:fig1}(d)} . Influenced by located magnetic field as shown as the black arrows, signals with same frequency and different amplitude can be detected in adjacent sensing channels (represented by boxes with white borders in \textbf{Figure \ref{fig:fig1}(d)} ). Due to lower electron spin between channels and higher-order photon angular momentum mode, pumping beam with space-variant polarization exhibits lower crosstalk compared with conventional uniformly polarized pumping scheme.

\subsection{Theoretical Analysis: Relaxation due to Polarization Inhomogeneity}

\begin{figure}[ht!]
\centering
  \includegraphics[width=0.8\linewidth]{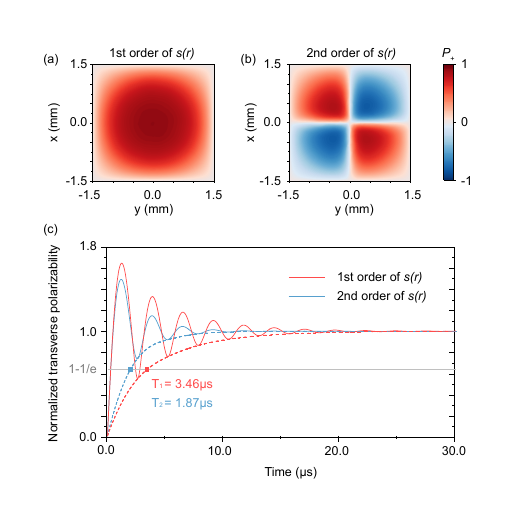}
  \caption{Steady-state distribution and transient response of transverse polarization. Fundamental mode of $P_+(\infty)$ induced by (a) $1^{st}$ and (b) $2^{nd}$ order of $s(r)$; (c) The transient response of overall transverse polarization induced by $1^{st}$ and (b) $2^{nd}$ order distribution of photon angular momentum.}
  \label{fig:fig2}
\end{figure}

While conventional theoretical models predominantly assume spatially homogeneous spin polarization distributions within atomic vapor cells, this work expanding the electronic spin polarization $P(r)$ and optical angular momentum field $s(r)$ using the eigenstates $u_i(r)$ of the Laplacian operator $\nabla^2$ to investigate the relaxation dynamics mediated by excitation of higher-order atomic polarization modes through photonic angular momentum modulation\cite{wang2024mode}.

\begin{equation}
  \textbf{P}(\textbf{r}) = \sum_{i=1} P^{(i)} u_i(\textbf{r}) 
\end{equation}

\begin{equation}
s(\textbf{r}) = \sum_{i=1} s^{(i)} u_i(\textbf{r})
\end{equation}

Substitute these into the Bloch equation yields:
\begin{equation}
\frac{d}{dt} \sum_i \textbf{P}^{(i)} u_i(\textbf{r}) = \frac{\gamma_e}{Q} \textbf{B} \times \sum_l \textbf{P}^{(i)} u_i(\textbf{r}) - \frac{R_{tot}}{Q} \sum_l P^{(i)} u_i(\textbf{r}) + \frac{R_{op}}{Q} \sum_i s^{(i)} u_i(\textbf{r}) \hat{z} + D \sum_i \textbf{P}^{(i)} \nabla^2 u_i(\textbf{r})
\end{equation}

Rewrite the above equation with the eigenvalue values of the $\nabla^2$:
\begin{equation}
\frac{d}{dt} \sum_i \textbf{P}^{(i)} u_i(\textbf{r}) = \frac{\gamma_e}{Q} \textbf{B} \times \sum_l \textbf{P}^{(i)} u_i(\textbf{r}) - \frac{R_{tot}}{Q} \sum_l P^{(i)} u_i(\textbf{r}) + \frac{R_{op}}{Q} \sum_i s^{(i)} u_i(\textbf{r}) \hat{z} + D \sum_i \textbf{P}^{(i)} k_i^2 u_i(\textbf{r})
\end{equation}

As derived from the equation above, if the angular momentum distribution of light $s$ is considered to have spatial variation while the intensity (corresponding to $R_{op}$) remains spatially uniform, different modes will exhibit no coherence with each other and evolve independently. So, the Bloch equation of each order is:
\begin{equation}
\frac{d}{dt}\textbf{P}^{(n)} = \frac{\gamma_e}{Q} \textbf{B} \times \textbf{P}^{(n)} - \frac{R_{tot}}{Q} \textbf{P}^{(n)} + \frac{R_{\text{op}}}{Q} s^{(n)} \hat{z}-D\textbf{P}^{(n)}k_n^2
\end{equation}

Let $B_+=B_x+iB_y,P_+=P_x+iP_y$. Decomposing the equation into transverse (x,y) and longitudinal (z) components, with the z-component assumed to take the pumping steady-state value, yields:
\begin{equation}
\quad-\frac{R_{\text{tot}}}{Q} P_z^{(n)} + \frac{R_{\text{op}}}{Q} s^{(n)} - DP_z^{(n)} k_n^2 = 0 
\end{equation}
\begin{equation}
\frac{d}{dt} P_+^{(n)} = i \left(P_+^{(n)} \omega_q - \frac{\gamma_e}{Q} P_z^{(n)} B_+ \right) - \frac{R_{\text{tot}}}{Q} P_+^{(n)} - D P_+^{(n)} k_n^2
\end{equation}

where $\omega_q=\gamma_eB_z/Q$ denote the Larmor precession frequency. Solving the above equation yields the steady-state values of the longitudinal (z-direction) atomic polarization modes at each order:
\begin{equation}
P_z^{(n)} = \frac{R_{op} T_n s^{(n)}}{Q}
\end{equation}

where $T_n$ represents the relaxation time for each order of polarization modes:
\begin{equation}
\frac{1}{T_n} = \frac{R_{tot}}{Q} + Dk_n^2.
\end{equation}

Substituting the transverse polarization modes' evolution equations into the system and simplifying yields:
\begin{equation}
\frac{d}{dt} P_{+}^{(n)}=(i\omega_q - \frac{1}{T_n}) P_{+}^{(n)} - i \frac{\gamma_e}{Q} B_{+} P_{z}^{(n)}
\end{equation}

The solution to this differential equation is:
\begin{equation}
P_+^{(n)}(t) = -i \frac{\gamma_e}{Q(i \omega_q - 1/T_n)} B_+ P_z^{(n)} \left( 1 - e^{i \omega_q t - t/T_n} \right)
\end{equation}

The equation above describes the temporal evolution of the transverse polarization modes of each order. The time-varying term comprises a rotating oscillatory term at the Larmor frequency ($\omega_q$) and an exponential decay term with the relaxation time of each order as the decay constant. Expressing the overall transverse polarization as a sum of the individual modes, we obtain:
\begin{equation}
P_+(t,r) = \sum_{n} P_+^{(n)}(t) \, u_n(r)
\end{equation}

\textbf{Figure \ref{fig:fig2}} shows the results of theoretical calculations of steady-state distribution and transient response of transverse polarization for the $3\times3\times3\ mm^3$  vapor cell with typical parameters $D=0.18\ cm^2s^-1{}$, $R_{op}=400s^{-1}$, $R_{rel}=400s^{-1}$ and $T=150 ^\circ C $.

The steady-state temporal error is:
\begin{equation}
    err(t,r) = -i e^{i\omega_qt} \sum_n \frac{\gamma_e}{Q(i\omega_q - 1/T_n)} B_+ P_z^{(n)} e^{-t/T_n} u_n(r)
\end{equation}

The total error becomes:
\begin{equation}
|err|^2 = \int err(t,\textbf{r}) \, err^*(t,\textbf{r}) \, d\textbf{r} = \sum_n \left| \frac{\gamma_e B_+}{|Q(i\omega_q - 1/T_n)} \right|^2 \left| P_z^{(n)} \right|^2 e^{-2t/T_n}
\end{equation}

Define the effective relaxation time $t_0$ as the time required for the total error to decay to $1/e$ of its maximum value:
\begin{equation}
\sum_n \left| \frac{\gamma_eB_+}{Q(i\omega_q - 1/T_n)} \right|^2 \left| P_z^{(n)} \right|^2 e^{-2t_0/T_n} = e^{-2} \sum_n \left| \frac{\gamma_eB_+}{Q(i\omega_q - 1/T_n)} \right|^2 \left| P_z^{(n)} \right|^2.
\end{equation}

Based on the equation derived above, the relaxation times of 1st-order and 2nd-order of $s(r)$ are calculated to be 3.46 $\mu s$ and 1.87 $\mu s$, respectively. This theoretical analysis reveals that PICI pumping scheme reduces relaxation time by 46$\%$ between channels compared to uniform pumping scheme, therefore reducing the atomic diffusion crosstalk between channels. 

\section{Experimental Section}

\subsection{Design, Fabrication, Characterization of the Metasurface}

\begin{figure}[hb!]
\centering
  \includegraphics[width=0.8\linewidth]{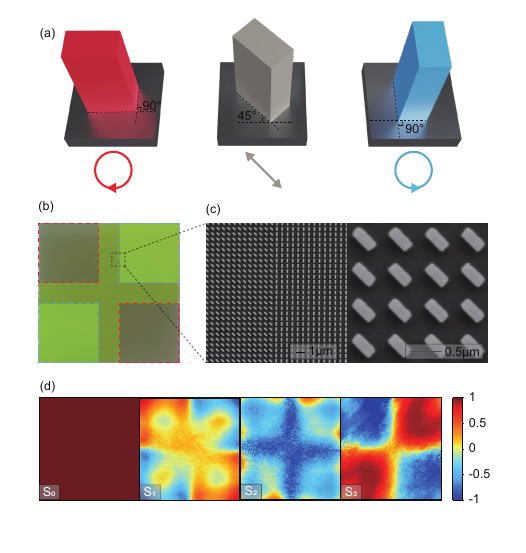}
  \caption{Design, characterization, and testing of space-variant polarization conversion metasurface. (a) Diagram of the meta-atom for three polarization transformers; (b) Optical micrographs and (c) SEM imaging of the space-variant polarization conversion metasurface; (d) Stokes parameter of the space-variant polarization conversion metasurface.}
  \label{fig:fig3}
\end{figure}

Over the past decade, the use of flat optics has been widely extended from wavefront shaping and focusing to more sophisticated manipulation of structured light, owing to rich meta-atom libraries, accurate full-wave simulations, and precise nanofabrication. Nowadays, a new generation of meta-optics can now perform parallel processing of the polarization of input light in the transverse plane, reducing the function of many polarizers and waveplates into a single optical component that can be integrated in on-chip magnetometers. The metasurface design methodology leverages the anisotropic optical response of nano-fins, where incident light propagating through each nanostructure acquires polarization-dependent phase delays and amplitude modulations along the major and minor axes, governed by the nano-fin's cross-sectional dimensions. Furthermore, rotating the nano-fin's in-plane orientation introduces a third degree of freedom that typically manifests as a Pancharatnam-Berry phase. 

Significant challenges in multi-parameter optimization must be overcome to accurately achieve polarization control in this PICI pumping scheme, which involves constructing large-scale parametric dataset through systematic variation of critical dimensions, followed by rigorous filtering to identify geometric configurations that satisfy the orthogonally aligned principal axes condition. Implementing this methodology demands substantial computational resources and often achieve limited design accuracy. Inverse-design and machine-learning techniques will be key tools for addressing this challenge by searching nonintuitive design spaces. In this study, a reverse optimization algorithm to directly optimize the structural dimensions of the nano-fin unit by using the DOP (Degree of Polarization) as the fitness function\cite{liang2024metasurface,liang2025space}. The detailed design methodology and process of nanofabrication are provided in the Supporting Information.

The meta-atoms of the three polarization transformers are schematically illustrated in \textbf{Figure \ref{fig:fig3}(a)}. The optical micrographs of the space-variant polarization conversion metasurface in \textbf{Figure \ref{fig:fig3}(b)} express very characteristic patterns. SEM imaging as shown in \textbf{Figure \ref{fig:fig3}(c)} displayed the interface of LP/RCP regions and the details of the meta-atoms in linear polarization region. Subsequently, a Stokes parameter measurement system, consisting of a quarter-wave plate (QWP), linear polarizer, and beam imaging system, is employed to systematically characterizes the performance of this space-variant polarizer\cite{schaefer2007measuring}. The experimental results displayed in \textbf{Figure \ref{fig:fig3}(d)} demonstrate that the transmitted light achieves a clear space-variant polarization distribution.

\subsection{Crosstalk Characterization via Located Fictitious Magnetic Field}

\begin{figure}[ht!]
\centering
  \includegraphics[width=0.8\linewidth]{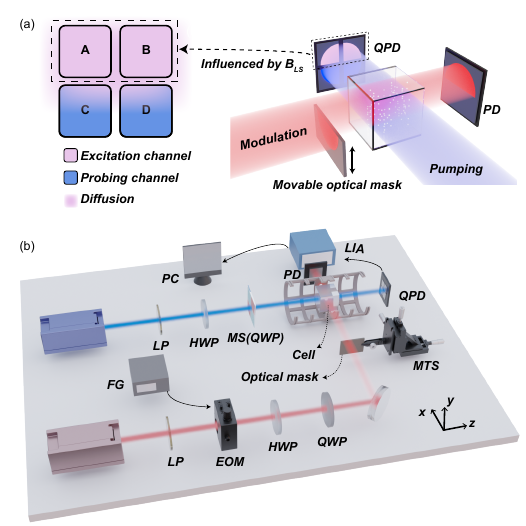}
  \caption{Experiment setup for crosstalk characterization through Optically Induced Fictitious Magnetic Field. (a) Diagram of localized excitation method: By modulating the incident light (red) and spatially confining it with an optical mask, a measurable fictitious magnetic field is generated within a localized region of the pump light (blue); (b) Experimental setup for characterization of crosstalk. MS, metasurface; FG, function generator; EOM, electro-optic modulator; MTS, manual linear translation stages; QPD, quadrant photodiode (with 100 $\mu m$ channel spacing); LIA, lock-in amplifier.}
  \label{fig:fig4}
\end{figure}

Critical to the comparison of crosstalk between the proposed PICI pumping scheme and the conventional QWP-based scheme is to accurately exciting a single channel with a spatially restricted magnetic field and then monitoring the crosstalk in the adjacent channels. However, the current coil-based magnetic field control systems in conventional magnetometers are ill-equipped to solve this challenge. Therefore, this research leveraging the power of localized fictitious magnetic field to selectively excite specific sub-regions in the atomic ensemble as shown in \textbf{Figure \ref{fig:fig4}(a)} and compares the amplitude ratio (channel A/channel C) between the two schemes based on metasurface and QWP. After passing through the vapor cell as shown in \textbf{Figure \ref{fig:fig4}(a)}, the intensity of the pumping beam is simultaneously detected by four independent channel photodetectors (quadrant photodiode detector, QPD), forming a four-channel atomic magnetometer sensitive to the spatial distribution of the magnetic field. At the same time, a detuned circularly polarized beam is incident the vapor in the direction perpendicular to the pumped light, covering the pumped light as completely as possible. Subsequently, controlled fictitious magnetic field is generated inside the vapor cell by modulating the intensity of the detuned circularly polarized light using an electro-optical modulator (EOM). Influenced by optical modulation, a response signal proportional to the fictitious magnetic field is captured by the photodetector. In addition, a movable optical mask is utilized to limit the range of influence of the modulated light on the pumping light, which realizes a spatially controlled fictitious magnetic field within the vapor cell. However, limited by millimeter spot size and isolated test environments, it is still very challenging to precisely control the fictitious magnetic field so that it only affects channel AB while detecting crosstalk on channel CD as shown in \textbf{Figure \ref{fig:fig4}(a)}. Thus, the entire displacement process of the optically masked, from fully masked to fully unmasked modulated light, is broken down into multiple test intervals. The ratio of signal amplitudes between channels A and C is recorded in each mask position as an indirect method of assessing crosstalk between the channels of the two pump schemes. The specific experimental setup and steps are detailed as follows.

The experimental setup is shown in \textbf{Figure \ref{fig:fig4}(b)}. A   borosilicate cubic cell containing a droplet of $^{87}Rb$ and $720 \ Torr$ of $N_2$ buffer gas serves as a sensitive source for magnetometers, heated to $433\ K$ using a twisted pair winding resistor driven by a 240 kHz AC electronic current\cite{zhou2025non}. A multilayer cylindrical magnetic shield composed of four layers of $\mu$-metal and an aluminum shield achieves a quasistatic shielding factor of at least $10^5$, guaranteeing a low-noise magnetic environment with a residual magnetic field measuring less than 5 nT. The outermost layer of aluminum shielding is designed to reduce high-frequency magnetic noise. Furthermore, a cylindrical uniform field coil is employed to control the triaxial magnetic field inside the cylindrical magnetic shield\cite{zhou2022,jia2025reducing,long2023situ}. A collimated 795 nm circularly polarized beam (diameter: 2.3 mm, power: 650 $\mu $W) propagating orthogonally to the primary pump laser (D1 line, $\sigma^+$, 1 mW) is frequency-detuned by +80 GHz from the Rb D1 transition to minimize absorption while maintaining sufficient light shift. The beam intensity is sinusoidally modulated at 30 Hz using an electro-optic modulator (Thorlabs, NEL03A/M), generating a time-varying fictitious magnetic field described by:
\begin{equation}
B_{\rm LS}(t) = B_0 \sin(60 \pi_m t), \quad B_0 = 3.5 n T.
\end{equation}

An optical mask on a manual linear stage (LBTEK, LS65-13) is utilized to spatially control the overlapping volume between the excitation and pumping beams, thereby confining the fictitious magnetic field into targeted submillimeter zones. The optical mask, translated vertically in 200 $\mu $m increments, spatially tunes the influence of the fictitious magnetic field—progressively shifting from no channel coverage to isolated activation of Channel A (at about 50$\%$ exposure) and finally to full-channel coverage (Channels A and C at 100$\%$ transmittance). 

Define the Position-Dependent Isolation Performance (PDIP) as

\begin{equation}
PDIP(x) = 20\log_{10} (\frac{|V_A(x)|}{|V_C(x)|  Q_c}
\end{equation}

Where $V_A(x)$ and $V_C(x)$ represents the complex amplitude of the 30 Hz response of the main channel A and the adjacent channel C at position x (obtained by Fourier transform). denotes the signal conversion efficiency of adjacent channel C (In order to eliminate the inherent errors between channels due to detector sensitivity, magnetometer scale factor, etc.).

The detailed experimental procedure is as follows:

1.	Initialize magnetometer in spin-exchange relaxation-free (SERF) regime\cite{seltzer2008developments} (T = 150$^\circ C$, $^{87}Rb$ density $n_{Rb}=1.01\times10^{14}cm^{-3}$ ); 

2.	Set optical mask to start position (x=0 mm, no channel interference) and acquire 60 s noise floor data;

3.	Perform stepwise translation (13 positions, x = 0 mm to x = 2.4 mm) with 200 $\mu m$ steps;

4.	At each position: 

Apply modulated fictitious field ($B_{LS}(t)=B_0\sin{60\pi_m t}$);
Simultaneously measure the 30 Hz signal amplitudes in the stimulated channel (A) and adjacent channel (C) using a synchronized dual-phase lock-in amplifier (Zurich Instruments MFLI), enabling crosstalk quantification at the fictitious magnetic field's modulation frequency.

\section{Results and Discussion}

\begin{figure}[ht!]
\centering
  \includegraphics[width=0.8\linewidth]{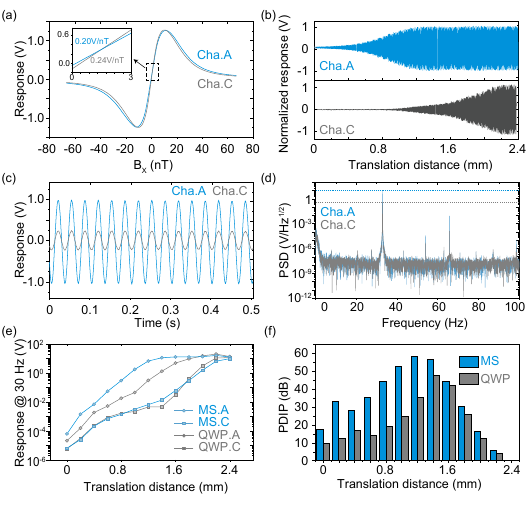}
  \caption{Experimental results for the characterization of crosstalk. (a) Dispersion-like magnetic response curves and scale factor of channel A and C; (b) Amplitude responses of Channels A and C during continuous optical mask translation (0 $\sim$ 2.4 mm range); (c) Time-domain and (d) frequency-domain amplitude responses of primary/adjacent channels under uniform circular polarization pumping (optical mask: x = 1.8 mm); (e) Amplitudes of 30 Hz signals (Channels A/C) under two pumping schemes in response to the location of optical mask; (f) Position-dependent isolation performance (PDIP, Channels A/C) under PICI scheme and uniform circular polarization pumping configurations.}
  \label{fig:fig5}
\end{figure}
Based on the experimental setup, a comparative evaluation of the response of magnetometer and crosstalk suppression between pumping mode based on space-variant polarization conversion metasurface and QWP is shown in \textbf{Figure \ref{fig:fig5}}. By scanning the magnetic field $B_x$ in SERF regime, the dispersion-like magnetic response curves and scale factors of channels A (0.2 $V\ nT^{-1}$) and C (0.24 $V\ nT^{-1}$) is demonstrated in \textbf{Figure \ref{fig:fig5}(a)}. The discrepancy in scale factors is likely originates from nonuniform light intensity distribution and variations in photoelectric conversion efficiency among different area of QPD. To address these systematic variations, signal conversion efficiency $Q$ is introduced into the calculation of the PDIP. This correction ensures accurate crosstalk quantification independent of inherent sensor responsivity differences. \textbf{Figure \ref{fig:fig5}(b)} tracks the amplitude responses of Channels A and C as the optical mask translates from full occlusion (x = 0) to complete exposure (x = 2.4 mm). Due to inherent alignment limitations in both pumping scheme, which make it challenge to precise operate optical mask to fully occlude Channel C while completely exposing Channel A, the entire displacement range is evenly discretized into 13 test intervals with 200 µm increments. For example, \textbf{Figure \ref{fig:fig5}(c)} indicates the time-domain amplitude response curves of the exciting channel and probing channel under uniform circularly polarized pumping with the optical mask positioned at x=1.8 mm. The corresponding frequency-domain analysis in \textbf{Figure \ref{fig:fig5}(d)}, obtained through Fourier transformation of the data in \textbf{Figure \ref{fig:fig5}(c)}, displays the power spectral density (PSD) profiles for both channels. The PDIP at this position is calculated from the 30 Hz components (corresponding to the modulation frequency of the fictitious magnetic field), where:

\begin{equation}
PDIP_{QWP}(1.8) = 20\log_{10} \left( \frac{|V_A(1.8)|}{V_C(1.8) | {Q}_C} \right) = 25.9\ dB
\end{equation}

This value quantifies the amplitude ratio between channels, serving as an evaluation for crosstalk in different pumping schemes. Following this methodology, \textbf{Figure \ref{fig:fig5}(e)} indicates the responses of the 30 Hz power spectral density (PSD) responses for channels A and C under schemes based on space-variant polarization conversion metasurface and QWP. Notably, in the PICI pumping scheme, signals in channel C exhibit a pronounced temporal delay relative to those in channel A, probably due to the establishment of distinct spatial boundaries caused by linearly polarized light absorption mechanisms. Furthermore, our repeated experiments discover that although the exposure area progressively increases with optical mask displacement, the response amplitude in each channel does not exhibit perfect monotonic increase. Peaks of amplitude exceeding adjacent position are likely attributed to mask-induced optical diffraction. \textbf{Figure \ref{fig:fig5}(f)} compares channel isolation performance between pumping schemes. It is observed that the metasurface based pumping scheme suppresses more crosstalk at each mask location. Compared with conventional uniformly polarized pumping scheme that achieves 20 dB average crosstalk ratio, the PICI pumping scheme demonstrates up to 32 dB average crosstalk ratio between adjacent sensing channels. This 12 dB average enhancement corresponding to a 74.9$\%$ average crosstalk reduction and promising three times spatial resolution improvement. In addition, the difference between the PDIPs of the two pumping modes reaches a maximum of 27.9 dB at 1 mm, corresponding to 94.7$\%$ crosstalk relative reduction. These results directly attributable to polarization gradients of the space-variant polarization conversion metasurface that enable isolation of atom polarization in single SERF ensemble.

\section{Conclusion}

Spin-exchange relaxation-free (SERF) atomic ensembles are traditionally understood to exhibit uniform time evolution due to rapid spin-exchange interactions that homogenize atomic states across the ensemble. In this study, we show that micrometer-scale modulation of the pumping polarization enables controlled divergence in the dynamics of spatially distinct sub-ensembles within a single SERF system, offering a new degree of independence previously unattainable. By combining the space-variant polarization metasurface with a thermal atomic ensemble in SERF regime, antiphase spin polarization is imposed to the atomic ensemble by the construction of calculated Zeeman sublevel populations, leading to the separation of distinct sensing channels. Theoretical modeling reveals that the polarization gradient introduced by the polarization-imposed channel isolation (PICI) pumping scheme reduces the relaxation distance by 46$\%$ compared with the uniform pumping scheme based on a quarter-wave plate (QWP). Proof-of-concept experiments utilizing position-related light-modulated fictitious magnetic fields (3.5 nT @ 30 Hz) achieved spatially selective excitation of vapor cell sub-regions with 200-$\mu m$ precision. Inter-channel crosstalk between the two pumping schemes based on proposed metasurface and traditional QWP is quantified by the amplitude ratio of adjacent channel (100 $\mu m$ spacing) responses to the fictitious magnetic field. Experimental results demonstrating up to 32 dB average crosstalk ratio between adjacent sensing channels (scale factor: 0.2 $V\ nT^{-1}$ for channel A and 0.24 $V\ nT^{-1}$ for channel C). Compared with conventional uniformly polarized pumping scheme that achieves 20 dB average crosstalk ratio, the scheme with atom polarization isolation demonstrating a 12 dB average enhancement in channel independence. This result corresponding to a 74.9$\%$ average crosstalk reduction and promising three times spatial resolution improvement. Future work will focus on the heterogeneous integration with VCSEL\cite{zhou2025non,zhou2022application} laser, integration with micro-fabricated vapor cell, and optimization spacing between channels, promoting the development of chip-scale SERF magnetometers to enable ultra-sensitive and high-resolution measurements of extremely weak biomagnetism fields.

\bibliography{references}
\bibliographystyle{unsrt}  



\end{document}